\documentclass[conference]{IEEEtran}
\usepackage{graphicx}
\usepackage{amsmath,amssymb}
\usepackage{cite}
\usepackage{booktabs}
\usepackage{float}
\usepackage[linesnumbered,vlined,ruled]{algorithm2e}
\usepackage{hyperref}
\usepackage{adjustbox}
\usepackage[justification=centering]{caption}

\def\BibTeX{{\rm B\kern-.05em{\sc i\kern-.025em b}\kern-.08em
    T\kern-.1667em\lower.7ex\hbox{E}\kern-.125emX}}

\begin{document}

\title{UAV-Assisted 6G Communication Networks for Railways: Technologies, Applications, and Challenges}

\author{
Aamer Mohamed Huroon\textsuperscript{*‡},
and Li-Chun Wang\textsuperscript{*}\\
\IEEEauthorblockA{
\textsuperscript{*}Department of Electrical and Computer Engineering, National Yang Ming Chiao Tung University \\
\textsuperscript{‡} Department of Electrical and Electronic Engineering, University of Nyala \\
Email: \{aamer.ee08,  wang\}@nycu.edu.tw}
}

\maketitle

\begin{abstract}

Unmanned Aerial Vehicles (UAVs) are crucial for advancing railway communication by offering reliable connectivity, adaptive coverage, and mobile edge services . This survey examines UAV-assisted approaches for 6G railway needs including ultra-reliable low-latency communication (URLLC) and integrated sensing and communication (ISAC). We cover railway channel models, reconfigurable intelligent surfaces (RIS), and UAV-assisted mobile edge computing (MEC). Key challenges include coexistence with existing systems, handover management, Doppler effect, and security. The roadmap suggests work on integrated communication-control systems and AI-driven optimization for intelligent railway networks.
\end{abstract}


\section{Introduction}
\IEEEPARstart{T}{he} rapid evolution toward intelligent railway systems demands unprecedented communication capabilities \cite{1, 2,3}. Sixth-generation (6G) wireless technologies are expected to support ultra-reliable low-latency communication (URLLC), massive connectivity, and integrated sensing and communication (ISAC) for next-generation railways \cite{4,5}. However, traditional terrestrial infrastructure faces challenges in providing seamless coverage, especially in remote areas and tunnels. Unmanned Aerial Vehicles (UAVs) have emerged as a promising solution, offering flexible deployment, adaptive coverage, and mobile edge computing capabilities \cite{6}.

The integration of UAVs into railway communication networks represents a paradigm shift from static to dynamic infrastructure. Unlike fixed base stations, UAVs can be dynamically positioned to provide coverage where and when needed, making them particularly valuable for railway applications where connectivity requirements change rapidly with train movement \cite{7}. This multifunctional capability aligns perfectly with the 6G vision of pervasive connectivity and intelligent network management \cite{8}.

This paper provides a comprehensive survey of UAV-assisted 6G communication networks for railways. Section II discusses communication architectures. Section III examines channel models and management strategies. Section IV analyzes RIS-aided schemes. Section V explores UAV-assisted MEC. Section VI addresses key challenges, while Section VII presents future research directions.

\section{UAV-Assisted Communication Architectures}
\subsection{Airborne Base Stations and Relays}
UAVs function as airborne base stations (ABS) to provide immediate connectivity where terrestrial infrastructure is unavailable \cite{9}. In railway scenarios, ABS are valuable for covering gaps in remote regions, during temporary events, or in emergencies. UAV mobility allows them to follow trains along routes, maintaining consistent signal strength throughout the journey \cite{10}.
\begin{figure*}[htbp]
    \centering
    \includegraphics[width=0.9\linewidth]{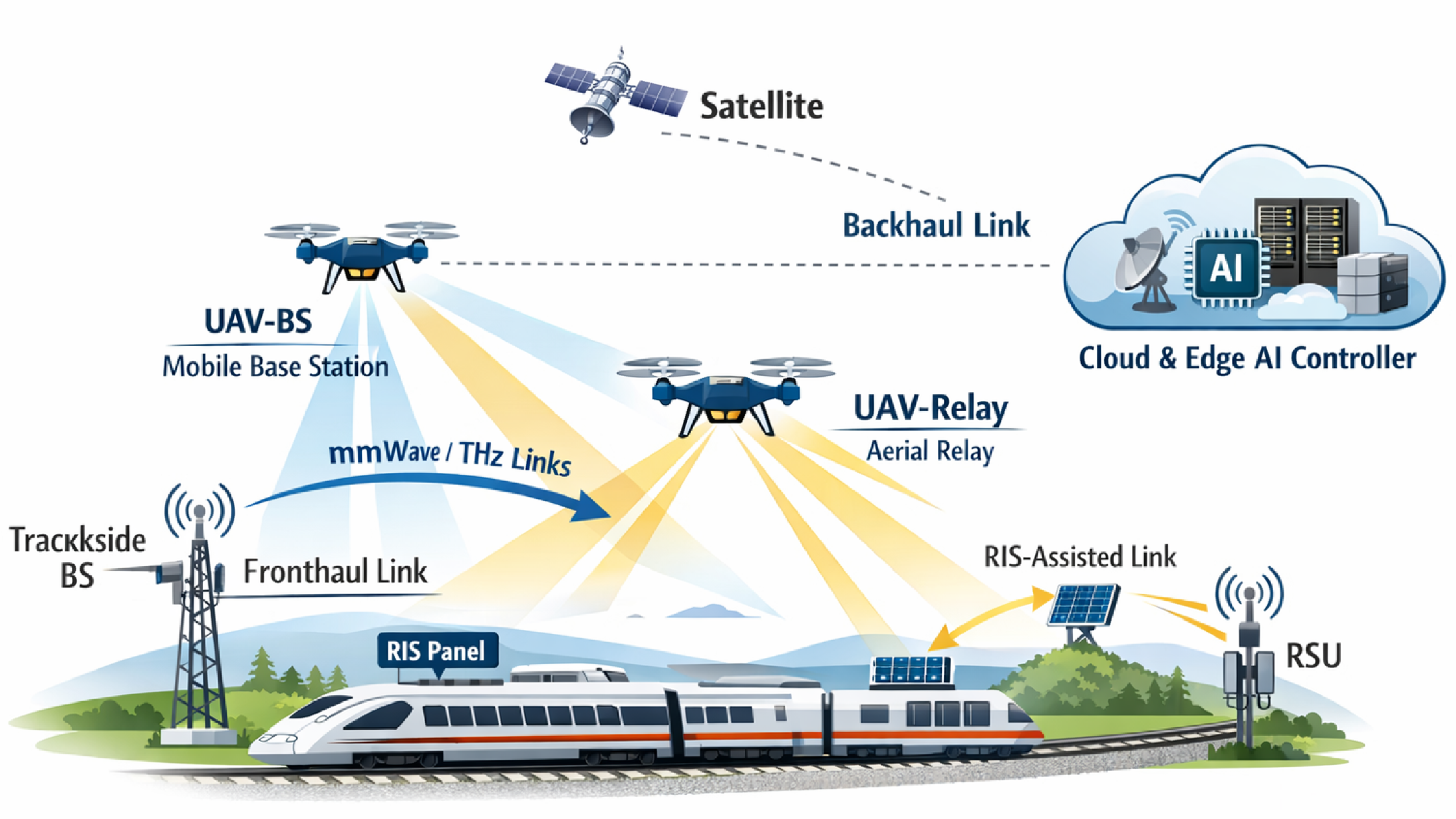}
    \caption{Hierarchical architecture of UAV-assisted 6G railway networks showing integration of UAVs, terrestrial base stations, RIS, and AI-driven control systems. The architecture demonstrates multi-layer connectivity with aerial, terrestrial, and satellite components.}
    \label{fig:architecture}
\end{figure*}
As relays, UAVs extend coverage of existing base stations by creating multi-hop links between terrestrial infrastructure and trains. This is crucial where direct links are obstructed by terrain or buildings. Strategic placement of UAV relays improves signal-to-noise ratio and reduces path loss in challenging environments \cite{11}.

\subsection{UAV-Enabled Integrated Sensing and Communication}
UAV-enabled integrated sensing and communication (ISAC) allows simultaneous collection of environmental data and transmission of communication signals \cite{12,13}. In railway contexts, this supports track monitoring, obstacle detection, security surveillance, and predictive maintenance while maintaining reliable communication links.
For sensing applications, UAVs equipped with cameras, LiDAR, radar, or other sensors perform detailed inspections of railway infrastructure. The real-time data can be processed locally or transmitted to central systems for analysis, enabling proactive maintenance \cite{14}.

\subsection{Network Architecture and Integration}
Integration of UAVs into existing railway networks requires careful architectural considerations. Two primary approaches have emerged: UAV-assisted and UAV-integrated networks. UAV-assisted networks use UAVs as supplementary elements without fundamentally altering existing architecture. UAV-integrated networks treat UAVs as intrinsic network elements, requiring more significant modifications but offering greater performance improvements \cite{15}.

\begin{figure*}[htbp]
    \centering
    \includegraphics[width=1.0\linewidth]{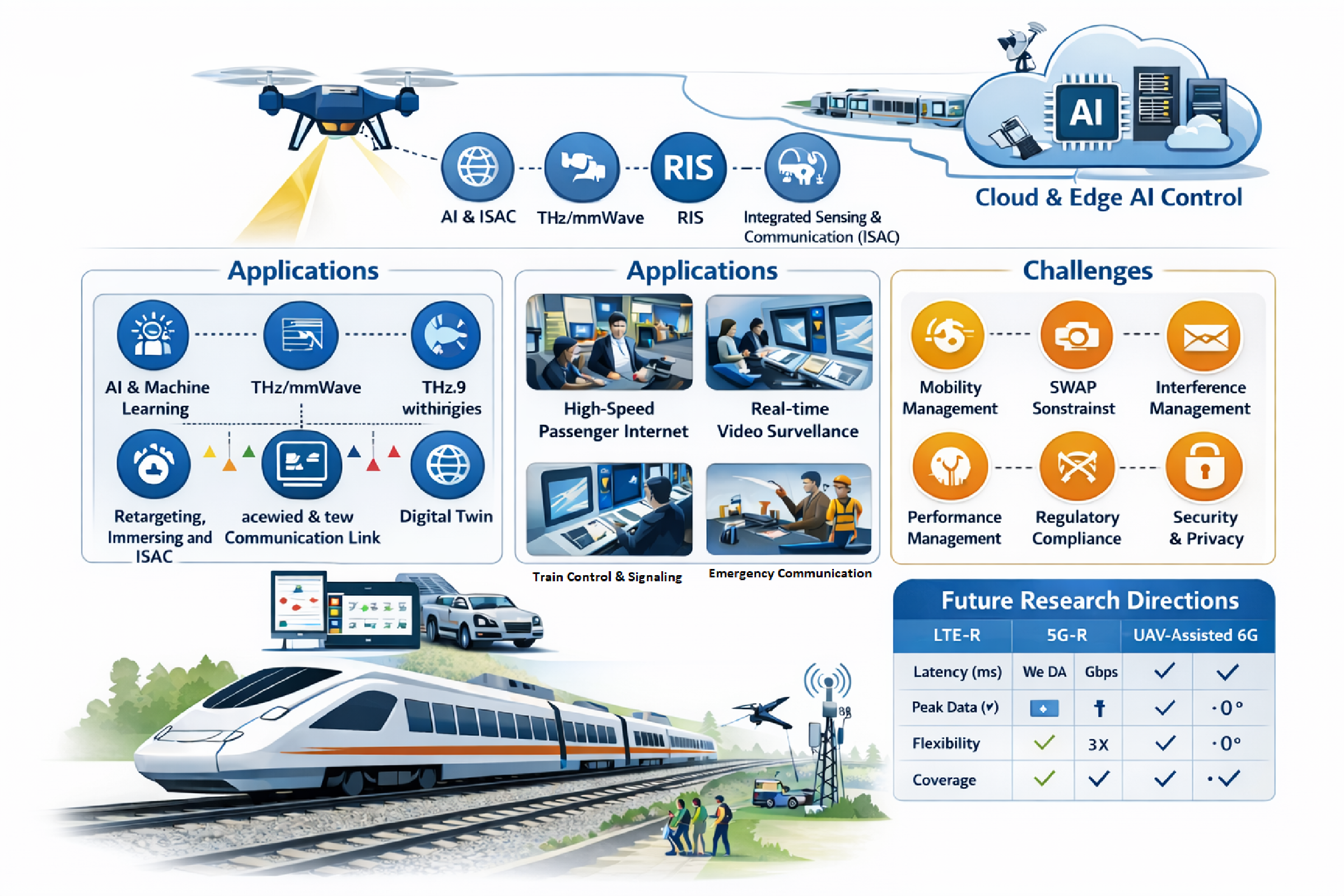}
    \caption{UAV-assisted 6G communication networks deployment strategies in railway environments showing different operational modes: airborne base stations following trains, fixed-wing UAVs for long-range coverage, and rotary-wing UAVs for precise positioning. The figure illustrates trajectory optimization and coverage patterns.}
    \label{fig:deployment}
\end{figure*}

A hierarchical architecture is often employed, consisting of aerial, terrestrial, and satellite layers. The aerial layer comprises UAVs at different altitudes, with high-altitude platforms (HAPs) providing wide-area coverage and low-altitude platforms (LAPs) offering localized support \cite{16}.

\section{Railway Channel Models and Management}
\subsection{Railway Channel Characteristics}
Accurate channel modeling is fundamental for UAV-assisted railway communication systems \cite{17}. Railway environments present unique propagation characteristics differing from typical urban scenarios. The linear topology of railway corridors creates distinctive propagation conditions, with signals often traveling parallel to tracks and experiencing reflections from nearby structures.
The air-to-ground (A2G) channel between UAVs and trains exhibits different characteristics compared to terrestrial links. A2G channels typically experience less scattering and shadowing, especially when UAVs operate at higher altitudes with clear line-of-sight paths. However, channel quality is significantly influenced by UAV altitude, elevation angle, and surrounding environment \cite{1}.

\subsection{Resource Management and Optimization}
Efficient resource management is critical for maximizing performance while meeting diverse quality of service requirements. Resources include spectrum, transmission power, computational resources, and UAV trajectories. The dynamic nature of railway environments necessitates adaptive resource allocation strategies \cite{18}.

Spectrum management faces challenges due to potential interference between UAV links and existing terrestrial systems. Dynamic spectrum sharing techniques, including cognitive radio approaches, can improve spectral efficiency \cite{19}. Power control affects both communication performance and UAV energy consumption, requiring optimization to maintain reliable connections while extending flight time \cite{20}.

\subsection{Energy Efficiency and Flight Time}
Energy efficiency is paramount as UAV flight time is limited by battery capacity. Various approaches improve energy efficiency, including optimized trajectory planning, energy-aware resource allocation, and energy harvesting techniques \cite{21}.

Trajectory optimization balances communication performance and energy consumption. By carefully planning flight paths, it's possible to maintain strong communication links while minimizing propulsion energy \cite{22}. Energy harvesting technologies like solar panels offer promising avenues for extending operational time.

\section{RIS-Aided Schemes for Performance Enhancement}
\subsection{Fundamentals of Reconfigurable Intelligent Surfaces}
Reconfigurable Intelligent Surfaces (RIS) have emerged as a transformative technology for 6G networks, offering unprecedented control over the propagation environment \cite{23,24}. An RIS consists of a planar array of metamaterial elements whose electromagnetic properties can be dynamically controlled. By adjusting phase, amplitude, or polarization of incident waves, RIS can constructively combine signals at intended receivers or destructively cancel interference \cite{25}.

Integration of RIS with UAVs creates synergistic benefits for railway communications. Mounting RIS on UAVs combines positioning flexibility with signal manipulation capabilities, enabling dynamic optimization of the propagation environment in three-dimensional space \cite{26}.

\subsection{RIS-Assisted UAV Communication for Railways}
RIS-assisted UAV communication offers distinct advantages. First, RIS can extend effective coverage by creating virtual line-of-sight paths when direct links are obstructed. Second, RIS can enhance signal strength at specific locations, such as inside train carriages, by focusing electromagnetic energy toward intended receivers \cite{28}.

Deployment strategies include UAV-mounted RIS, ground-deployed RIS, and hybrid architectures. UAV-mounted RIS offers maximum flexibility as position and orientation can be dynamically adjusted. Ground-deployed RIS provides more stable platforms but with less adaptability.

\section{UAV-Assisted Mobile Edge Computing}
\subsection{Edge Computing Architecture}
Mobile Edge Computing (MEC) brings computational resources closer to end users, reducing latency and saving backhaul bandwidth \cite{29}. UAV-assisted MEC extends this by deploying edge computing capabilities on aerial platforms, creating dynamic computing infrastructure that can follow trains \cite{30}.
The hierarchical edge computing architecture typically consists of three tiers: cloud data centers, terrestrial edge nodes, and aerial edge nodes on UAVs. Tasks can be dynamically allocated across these tiers based on latency requirements, computational demands, and data sensitivity.

\subsection{Computation Offloading and Task Allocation}
Computation offloading allows resource-constrained devices to transfer computationally intensive tasks to more powerful edge servers. In UAV-assisted railway networks, offloading can occur from trains to UAVs, from UAVs to terrestrial edge nodes or cloud, and between collaborating UAVs \cite{31}.

The offloading decision process considers task characteristics, current system state, and economic considerations. Mathematical models often formulate the problem as constrained optimization aiming to minimize latency, energy consumption, or weighted combinations \cite{32}.

\subsection{Application Scenarios}
UAV-assisted MEC enables numerous applications enhancing safety, efficiency, and passenger experience. For safety and security, real-time video analytics process surveillance footage to detect threats. Predictive maintenance applications analyze sensor data to identify early signs of equipment failure \cite{33}. Passenger services include augmented reality interfaces, real-time translation, and immersive entertainment systems.

\section{Key Challenges and Solutions}
\subsection{Coexistence with Existing Systems}
Integration presents significant coexistence challenges with existing railway communication systems like GSM-R and LTE-R. UAV-assisted networks must operate alongside these without causing harmful interference \cite{34}.
Spectrum sharing represents a primary technical challenge. Dynamic spectrum access techniques enable UAVs to identify and utilize available spectrum while avoiding interference. Database-assisted spectrum sharing offers a complementary approach.

\subsection{Handover Management and Doppler Effect}
High train velocity creates frequent handovers and significant Doppler frequency shift \cite{35}. Handover management is more complex due to three-dimensional UAV mobility. Machine learning approaches can enhance handover decisions by predicting trajectory patterns \cite{36}.
The Doppler effect causes frequency shifts degrading communication performance. Compensation techniques include estimation and correction at receiver, pre-compensation at the transmitter, and robust signal design. Beyond mobility-related challenges, UAV-assisted railway networks face significant cross-layer design issues due to the tight coupling of communication, sensing, and computing functions. Physical- and MAC-layer decisions, such as beamforming and resource allocation, directly influence sensing accuracy and edge computing latency. These interactions are particularly critical for safety-sensitive railway applications, where unreliable or delayed information can have serious consequences. As a result, holistic cross-layer optimization frameworks are needed to jointly address reliability, sensing performance, and computational constraints. Additionally, scalability becomes challenging as the number of UAVs and connected devices grows along extended railway corridors. Coordinated multi-UAV operation and distributed control are therefore essential to maintain efficiency and system stability.

\subsection{Regulatory and Security Challenges}
Deployment must navigate complex regulatory landscapes governing aviation and telecommunications. Security challenges include vulnerability to eavesdropping, jamming, spoofing, and denial-of-service attacks \cite{37}. Physical security of UAVs is also a concern.
Privacy considerations require technical and policy measures including anonymization techniques, data minimization principles, and transparent privacy policies.

\section{Future Research Directions}
\subsection{Integrated Communication and Control Systems}
Future research should focus on tight integration of communication and control systems. This integrated approach is particularly important for autonomous train operations. Co-design involves developing mathematical frameworks that jointly optimize communication parameters and control policies \cite{22}.

Testbeds with actual UAVs, railway equipment, and communication infrastructure are essential for validating integrated designs under realistic conditions.

\subsection{Advanced Channel Modeling and Measurement}
Accurate channel characterization in UAV-assisted railway environments remains challenging. Future research should develop comprehensive models capturing three-dimensional UAV mobility, high train velocity, and specific propagation environments \cite{17}.
Measurement campaigns specifically targeting UAV-assisted railway scenarios are needed to validate and refine models. Machine learning techniques offer promising avenues for advancing channel modeling beyond traditional approaches.

\subsection{AI-Driven Optimization and Management}
Artificial intelligence will play an increasingly important role in optimizing and managing UAV-assisted railway networks. Reinforcement learning represents a particularly promising approach for managing complex control problems \cite{14}.
However, AI-driven approaches face challenges in safety-critical applications. Techniques for explainable AI and verifiable learning are needed to address concerns about model transparency and reliability. Moreover, AI-driven optimization enables proactive network management by predicting traffic demand, channel dynamics, and failures from real-time sensing data, thereby improving network reliability and resilience. Incorporating domain knowledge and safety-aware hybrid learning frameworks is essential to ensure stable and trustworthy operation in safety-critical railway systems.

\section{Conclusion}
UAV-assisted communication networks represent a promising paradigm for addressing stringent requirements of 6G railway systems. By providing flexible, adaptive coverage and enabling new services, UAVs can significantly enhance performance, reliability, and functionality \cite{1,6}.
Critical enabling technologies include airborne base stations, integrated sensing and communication, reconfigurable intelligent surfaces, and mobile edge computing. Each must be adapted to railway-specific constraints including high mobility, safety-critical operations, and linear coverage areas.
Important challenges remain in coexistence with existing systems, handover management, Doppler compensation, regulatory compliance, and security. Future research should focus on integrated communication-control systems, advanced channel modeling, and AI-driven optimization.

As 6G standards develop and UAV technologies mature, increasing adoption of UAV-assisted solutions is anticipated. The roadmap outlined provides guidance for researchers and practitioners working toward safer, more efficient, and more capable railway systems.

\section*{Acknowledgment}
This work has been partially funded by the National Science and Technology Council under the Grants 114-2811-E-A49-516-MY3, and 114-2221-E-A49 -185 -MY3, and 113-2218-E-A49 -027 -, and 114-2224-E-A49 -002 -, and 114-2218-E-A49 -019 -Taiwan. 
This work was supported by the Higher Education Sprout Project of the National Yang Ming Chiao Tung University and Ministry of Education (MOE), Taiwan.


\end{document}